\documentclass[final,5p,times,twocolumn]{elsarticle}

\usepackage{amssymb}
\usepackage{amsmath}
\usepackage{graphicx}
\usepackage{dcolumn}
\usepackage{bm}
\usepackage{array}%
\usepackage{tablefootnote}
\usepackage{booktabs}%
\usepackage{caption}
\usepackage{threeparttable}
\usepackage{multirow}
\usepackage{float}
\usepackage[T1]{fontenc}

\begin{document}
\begin{frontmatter}

\title{Extended R-matrix description of two-proton radioactivity}

\author[1,2,4]{Zhaozhan Zhang}
\author[1]{Cenxi Yuan\corref{cor1}}
\ead{yuancx@mail.sysu.edu.cn}
\author[3]{Chong Qi}
\author[1]{Boshuai Cai}
\author[4,5,6]{Xinxing Xu}

\address[1]{Sino-French Institute of Nuclear Engineering and Technology, Sun Yat-Sen University, Zhuhai 519082, China}
\address[2]{Department of Physics, Graduate School of Science, The University of Tokyo, Tokyo 113-0033, Japan}

\address[3]{KTH Royal Institute of Technology, Stockholm, SE-10691, Sweden}
      
\address[4]{CAS Key Laboratory of High Precision Nuclear Spectroscopy, Institute of Modern Physics, Chinese Academy of Sciences, Lanzhou, 730000, China}
\address[5]{School of Nuclear Science and Technology, University of Chinese Academy of Sciences, Beijing, 100049, China}
\address[6]{Advanced Energy Science and Technology Guangdong Laboratory, Huizhou, 516003, China}       

\cortext[cor1]{Corresponding author}

\begin{abstract}
Two-proton ($2p$) radioactivity provides fundamental knowledge on the three-body decay mechanism and the residual nuclear interaction. In this work, we propose decay width formulae in the extended R-matrix framework for different decay mechanisms, including sequential $2p$ decay, diproton decay, tri-body decay, and sequential two-diproton decay. The diproton and tri-body formulae, combined with information on the two-nucleon transfer amplitude and Wigner single-particle reduced width, can reproduce well experimental $2p$ radioactivity half-lives. For the case of $^{67}$Kr, theoretical predictions for direct $2p$ decay give much larger half-lives than the recent measurement from RIKEN. A combination of direct and sequential $2p$ emission is analyzed by considering a small negative one-proton separation energy and a possible enhanced contribution from the $p$-wave component. The present method predicts that $^{71}$Sr and $^{74}$Zr may be the most promising candidates for future study on $2p$ radioactivity. Our model gives an upper limit of 55(4) keV for the decay width of $4p$ emission in recently found four-proton resonant nuclide, $^{18}$Mg, which agrees with the observed width of 115(100) keV.
\end{abstract}


\end{frontmatter}


\section{Introduction}
Fruitful nuclear structure information can be revealed by radioactive decay studies. What is of particular interest now is extreme proton- or neutron-rich nuclei close and beyond the drip lines that may result in exotic decay. An exotic decay mode beyond the proton drip line is the two-proton ($2p$) radioactivity. It was predicted by Goldanskii \cite{goldansky1961two} and was first discovered in the decay of $^{45}$Fe 20 years ago \cite{pfutzner2002first}. A review of recent progress can be found in Ref. \cite{zhou2022}. The $2p$ radioactivity occurs in even-Z nuclei where one proton emission is energetically forbidden while $2p$ emission is allowed. It is often called true or direct $2p$ decay to distinguish from the sequential $2p$ decay, which can happen when the intermediate levels resulting from the $1p$ decay are accessible for another $1p$ emission. 

Schematically, the direct $2p$ emission can be described by two extreme scenarios: i) the emission of two completely uncorrelated protons, also called the tri-body decay. The simultaneously emitted two protons distribute isotropically but usually have similar energy; ii) the strongly correlated emission, also called the diproton emission. In this case, the decay undergoes through the penetration of a preformed ‘$^2$He’-cluster. The emitted two protons have a small angular distribution and share the same energy. Until now, only five nuclei have been observed that can have $2p$ radioactivity from the ground state ($^{45}$Fe \cite{giovinazzo2002two,pfutzner2002first}, $^{54}$Zn \cite{blank2005first}, $^{48}$Ni \cite{dossat2005two}, $^{19}$Mg \cite{mukha2007observation}, and $^{67}$Kr \cite{goigoux2016two}). A systematic study of $2p$ radioactivity with density functional theory suggests that this decay mode may not be limited to the light- and medium-mass regions but a typical feature for even-Z proton-unbound isotopes, and should be observed for nuclei in the heavier-mass region \cite{olsen2013landscape}.

The $2p$ emission characteristics are sensitive to detailed decay patterns, requiring theoretical methods to predict the $2p$ decay width with reasonable accuracy. The three-body model developed by Grigorenko et al. \cite{grigorenko2000theory,grigorenko2001two} contains rich ingredients for describing the two-proton decay process. In their formalism, the nuclear structure is considered in a simplified way, where only one single-orbital wave function is considered in the internal region, which is not realistic from the configuration mixing view. On the other hand, the R-matrix approach has been applied to study the $2p$ emission in $^{12}$O \cite{barker1999width,barker200112}, $^{45}$Fe \cite{brown2003di} and $^{18}$Ne \cite{brown2002di}. However, its application was limited due to the simplified treatment of the decay process.

In the present letter, we propose $2p$ decay width formulae for different decay mechanisms in the extended R-matrix framework to analyze some observed $2p$ radioactivity cases. In addition, we apply the model to study the possible heavier candidates for $2p$ radioactivity and discuss a recently observed $4p$ emitter $^{18}$Mg.

\section{Extended R-matrix theory for two-proton emission and beyond}
The R-matrix theory provides a basic framework for two-body decay and has been extended to discuss some specific situations of three-body decay, where the three-body disintegration can be treated as a succession of two two-body disintegrations \cite{lane1958r}. Here, we propose a realistic three-body extension by reviewing previous treatments and extend the discussion to a $4p$ emission case.

The standard form of the total decay width $\Gamma=\sum_c 2\gamma_c^2P_c$ originates from a fundamental relation in R-matrix theory,
\begin{equation} \label{eq1}
(E_2-E_1)\int{\Phi_2^* \Phi_1 d\tau}=\sum_{c}(V_{ 2c}^* D_{1c}-V_{1c}D_{ 2c}^*).
\end{equation}
The right-hand term results from using Green's theorem to transform the integral over the internal region into an integral over the surface. The wave function and its radial derivative at the surface can then be decomposed by the channel basis labeled by $c$, with $V_c$ and $D_c$ the respective overlap amplitudes. The channel basis involves two bound systems labeled by their spins and relative movement for the two-body decay. If one of the subsystems is unbound, the complete quantum numbers should include the internal energy $E$ and the subsequent open channel $r$ of the unbound system. The channel basis can then be labeled by $\widetilde{c}=\{c,E,r\}$ and Eq.~(\ref{eq1}) comes into,
\begin{equation}\label{eq2}
(E_2-E_1)\int{\Phi_2^* \Phi_1 d\tau}=\sum\limits_{\widetilde{c}}(V_{2\widetilde{c}}^* D_{1\widetilde{c}}-V_{1\widetilde{c}}D_{2\widetilde{c}}^*).
\end{equation}
The energy dependence of the overlap integral $V_{\widetilde{c}}$ or $D_{\widetilde{c}}$ can be approximated by \cite{lane1958r},
\begin{equation} \label{eq3}
{A_r(E)}^2=\frac{\Gamma_r \pi /2}{(E_r+\Delta_r-E)^2+\frac{1}{4}\Gamma_r^2},
\end{equation}
where $E_r$ is the resonance energy, $\Delta_r$ is the level shift expressed by $-\gamma_r^2 S_r(E)$, and $\Gamma_r$ is the partial decay width expressed by $2\gamma_r^2P_r(E)$. $P_r$ and $S_r$ are the Coulomb functions typically used in R-matrix theory. Finally, the extension to the three-body decay problem can be carried through the R-matrix theory with a modification that $\sum_{c}$ is everywhere replaced by $\sum_{\widetilde{c}} =\sum_{c,r}\int dE $. The total decay width of the three-body decay can then be expressed by,
\begin{eqnarray} \label{eq4}
\Gamma(Q_{2p})&=&\sum_{\widetilde{c}} 2\gamma_{\widetilde{c}}^2P_{\widetilde{c}}\nonumber \\
&=&\sum_{c,r}\int_0^{Q_{2p}}{dE 2\gamma_c^2 A_r(E)^2 P_c(Q_{2p}-E)}.
\end{eqnarray}
Eq.~(\ref{eq4}) was used to describe the diproton \cite{barker200112,brown2003di,brown2002di} and sequential $2p$ \cite{barker1999width} emission. For the diproton case, the energy amplitude $A_r(E)$ was formulated in terms of $p+p$ $s$-wave phase shift $\delta(E)$ with an analytical expression \cite{barker200112}. Compared with the three-body model by Grigorenko et al., Eq.~(\ref{eq4}) can consistently treat the structure information from the basic assumption of a mixing-state wave function and the resulting definition of the reduced width,
\begin{equation} \label{eq5}
\gamma^2=\theta^2\gamma_{sp}^2,
\end{equation}
where $\theta^2$ is the spectroscopic factor reflecting the weight of different configurations in the mixing-state wave function and $\gamma_{sp}^2$ is the single-particle reduced width.

Although Eq.~(\ref{eq4}) can consider the nuclear structure reasonably well, it treats the three-body disintegration in a simplified way and fails to describe the tri-body decay. Here, we propose a realistic three-body extension for the tri-body decay mechanism. In this case, the simultaneously emitted two protons are largely spatially separated and proton-proton interaction can be neglected compared to proton-core interaction. With the heavy core assumption, the Hamiltonian can be written as,
\begin{equation} \label{eq6}
H=H_{core}+T_X+T_Y+V_{core-X}+V_{core-Y}+\Delta V,
\end{equation}
where $X$ and $Y$ represent the single-particle states of the two emitted protons and $\Delta V$ is the residual interaction that can be neglected in the outer region. By applying the same operations in Eq.~(\ref{eq1}), we have,
\begin{align}\label{eq7}
(E_2-E_1)\int{\Phi_2^* \Phi_1 d\tau}=\sum_{i\in \{X,Y\}} \sum_{\widetilde{c}}(V_{2\widetilde{c}i}^* D_{1\widetilde{c}i}-V_{1\widetilde{c}i}D_{2\widetilde{c}i}^*).
\end{align}
The sums in $X$ and $Y$ result from Green's theorem to $T_{X}$ and $T_{Y}$, which moves the wave function to the surface and transforms it into one single-particle state with a resonance subsystem. Therefore, the channel basis can be labeled by $\widetilde{c}_X=\{c_X, E_Y, r_Y\}$ or $\widetilde{c}_Y=\{c_Y, E_X, r_X\}$. The tri-body decay width can then be approximated by replacing $\sum_c$ to $\sum_{\tilde{c}_i}$,
\begin{equation} \label{eq8}
\Gamma(Q_{2p})=\sum_{\mbox{\tiny$\begin{array}{c}
  i,j \in \{X,Y\}\\
  i \neq j\\ \end{array}$}}{\sum_{c,r}}\int_{0}^{Q_{2p}}{2\gamma_{ci}^2}{A_{rj}(E_j)^2 P_{ci}(E_i)dE_X,}
\end{equation}
with  $E_X+E_Y=Q_{2p}.$ The two protons are assumed to occupy paired states. Therefore, the angular momentum, the resonance energy, and the spectroscopic factor are the same for the two single-particle proton states. Compared with the diproton formula, Eq.~(\ref{eq8}) shows explicit dependence on the properties of the resonance core+$p$ subsystem.   

Other multi-body decay can follow similar decay mechanisms and thus be described by the above formulae with slight modifications. For instance, two recently reported $4p$ emitters ($^8$C \cite{charity20102} and $^{18}$Mg \cite{jin2021first}) are observed to decay in two sequential steps of direct $2p$ emission. We can then employ Eq.~(\ref{eq4}) by considering a sequential two-diproton emission. In this case, the internal energies of the emitted diprotons are neglected, and the decay width may be overestimated. To take into account this effect, we proposed a decay width formula in the form of,
\begin{eqnarray} \label{eq9}
\Gamma(Q_{4p})=\iiint A_1(E_1)^2 A_2(E_2)^2 A_3(E_3)^2\nonumber \\
\times 2\gamma_{l}^2P_l(E_{res})dE_1dE_2dE_3,
\end{eqnarray}
with $E_1+E_2+E_3+E_{res}=Q_{4p}$, where $E_1$ and $E_2$ are the internal energies of the first and second emitted diprotons, respectively. $E_3$ is the penetration energy of the second emitted diproton in the intermediate nucleus and $A_{1,2,3}$ are the corresponding energy amplitudes. Detailed discussion will be presented in Section~\ref{sec:level4}.

In the following application, the single-particle reduced width is calculated by Wigner width $3\hbar^2/2M_c R_c^2$ \cite{teichmann1952sum}, where $R_c$ is the channel radius defined by $1.45(A_1^{1/3}+A_1^{1/3})$. The spectroscopic factor $\theta_{2p}^2$ for finding two protons in the paired state is calculated by \cite{anyas1974nuclear},
\begin{equation} \label{eq10}
\theta_{2p}^2=(\frac{A}{A-2})^\lambda \times G^2\times TNA(S=0)^2,
\end{equation}
where $A$ is the mass of the parent nucleus, $G$ is the overlap of the two-proton shell-model wave function with the two-proton cluster wave function, and TNA is the two-nucleon transfer amplitude in paired states.

To conclude, in this section, we propose a realistic three-body extension to describe the tri-body decay and a sequential two-body extension to describe the sequential $4p$ decay. The decay widths in the R-matrix framework consider nuclear structure information reasonably well via spectroscopic factors and can combine with the configuration-interaction calculations to make the prediction.

\section{Results and Discussion}
\subsection{\label{sec:level1}Predicted half-life for observed $2p$ radioactivity}
The first case of the ground-state $2p$ radioactivity was identified in $^{45}$Fe by two experiments at GANIL \cite{giovinazzo2002two} and GSI \cite{pfutzner2002first}. These experiments were soon confirmed by an improved GANIL experiment with a similar set-up \cite{dossat2005two} and an experiment at Michigan State University with a time projection chamber (TPC) \cite{audirac2012direct}. Later, the same techniques with silicon detector and TPC were applied to observe the $2p$ events in $^{54}$Zn \cite{blank2005first} and $^{48}$Ni \cite{dossat2005two}. The $2p$ decay in $^{19}$Mg was observed with different techniques at GSI due to its much shorter half-life \cite{mukha2007observation}. Finally, the very recent case of $^{67}$Kr was reported in an experiment at RIKEN in Japan with a silicon detector telescope \cite{goigoux2016two}. 

The experimental data of the ground state $2p$ radioactivity are summarized in Table 1, including the observed $2p$ half-lives and decay energies for different $2p$ emitters. The single proton separation energies, except for that of $^{19}$Mg from experiment data \cite{mukha2007observation}, are obtained from theoretical predictions by calculating the Coulomb displacement energies to the known experimental masses of their mirror systems. These quantities provide necessary inputs for theoretical predictions.

Diproton and tri-body decay widths are calculated according to Eq.~(\ref{eq4}) with Barker's formulation \cite{barker200112} and Eq.~(\ref{eq8}), respectively, to analyze the ground state $2p$ radioactivity through different decay mechanisms. In the application, experimental $2p$ decay energies are used and the $\theta_{2p}^2$ values are determined with Eq.~(\ref{eq10}), in which the TNAs can be deduced from the configuration-interaction method. Details of the TNA calculation, including the choice of specific model spaces and effective interactions for different nuclei, were discussed in Ref.\cite{brown2019hybrid}. Results of the calculated $\theta_{2p}^2$ are listed in Appendix A.1, where a dominant paired state but with small purity is observed for each $2p$ emitter. Finally, the calculated $2p$ half-lives are compared with experimental data and different theoretical predictions through the three-body model \cite{grigorenko2003prospective, grigorenko2003two} and its shell model correction hybrid model \cite{brown2019hybrid}. The results are presented in FIG. 1 and Table 1.  

In the three-body model, a one-orbital wave function is expected and the result is sensitive to the angular momentum as shown in Appendix A.2. The lower angular momentum results can fit experimental data well and are chosen for comparison in this work. In addition, the three-body model makes no assumption on the decay mechanism. Both (i.e., diproton and tri-body) or even more complicated decay mechanisms are present simultaneously in the calculation. The distinction between them requires further correlation study. In this work, we have classified the simultaneous decay modes into two simplified pictures and studied them separately based on different assumptions. The calculated half-life considers the contribution from different configurations. The hybrid model has attempted to put more structure information in the three-body model by multiplying the single-orbital two-proton decay widths with the transfer amplitudes coherently or incoherently \cite{brown2019hybrid}. This treatment is different from the R-matrix formalism as shown in Eq.~(\ref{eq8}), where the TNA contributes to the decay width through both the reduced width and the corresponding energy amplitude. The dependence can not simply reduce to a coherent or incoherent form. Despite all these differences, they are comparable in the half-life prediction as shown in FIG. 1.

\begin{figure}[h]
\vspace{-0.4cm}
\setlength{\abovecaptionskip}{0cm}
\setlength{\belowcaptionskip}{-0.2cm}
\includegraphics[width=0.5\textwidth]{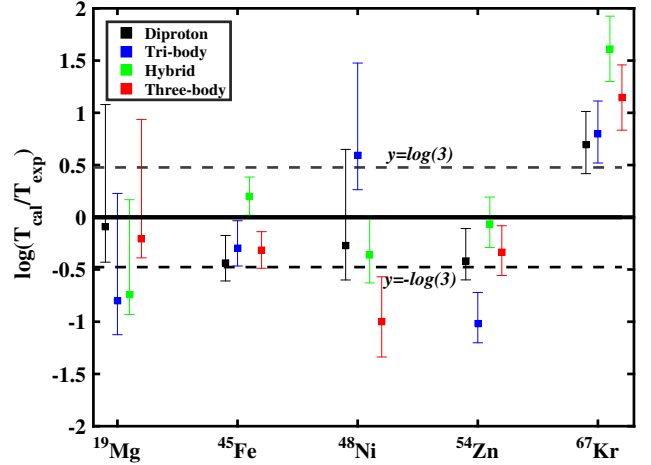}
\captionsetup{font={small}}
\caption{\label{fig:1} Ratios of experimental two-proton half-lives to different theoretical predictions, including the diproton formula of Eq.~(\ref{eq4}) with Barker's formulation \cite{barker200112}, the tri-body formula of Eq.~(\ref{eq8}), the three-body model \cite{grigorenko2003prospective, grigorenko2003two}, and its shell model correction Hybrid model in the incoherent form with $s^2$ contribution \cite{brown2019hybrid}. The values between two horizontal dash lines are different from experimental data within a factor of three.}
\end{figure}

From FIG. 1, except for the case of $^{67}$Kr, the diproton and hybrid models can well reproduce the experimental data within a factor of three. The results can be improved with a more precise measurement of $2p$ decay energies as the large $Q_{2p}$ sensitivity. The comparable tri-body half-lives may indicate a combined decay mechanism, which is consistent with previous experiments, where the $p-p$ angular correlations measurements showed a more complicated decay mechanism than a pure diproton case \cite{mukha2008proton, ascher2011direct, miernik2007two}. 
\begin{table*}[!ht]
\footnotesize
\setlength{\abovecaptionskip}{0cm} 
\setlength{\belowcaptionskip}{-0.2cm}
\captionsetup{font={normalsize}}
\caption{Two-proton half-lives in ms (except for $^{19}$Mg in ps) with different theoretical predictions, including the diproton formula of Eq.~(\ref{eq4}) with Barker's formulation \cite{barker200112}, the tri-body formula of Eq.~(\ref{eq8}), the three-body model \cite{grigorenko2003prospective, grigorenko2003two}, and its shell model correction Hybrid model using the incoherent form with $s^2$ contribution \cite{brown2019hybrid}.}
\resizebox{\textwidth}{!}{
\begin{tabular}{ccccccccc}
\toprule
 Nucleus& References&$S_p$(keV) &$Q_{2p}$(keV) &Expt. $T_{1/2}^{2p}$  &Diproton &Tri-body &Hybrid model &Three-body model\\ 
\midrule
\specialrule{0em}{1pt}{1pt}
$^{19}$Mg  &\cite{mukha2007observation} &550 &750(50) & 4.0(15)  & $3.26_{-2.25}^{+8.68}$  & $0.64_{-0.42}^{+1.49}$  & $0.73_{-0.17}^{+1.5}$  & $2.5_{-0.5}^{+6.5}$  \\
\specialrule{0em}{1pt}{1pt}
$^{45}$Fe  &\cite{miernik2007two} &24\tnote{$^\text{a}$}  &1154(16) & 3.7(4)  & $1.34_{-0.50}^{+0.81}$  & $1.86_{-0.69}^{+1.12}$  & 5.9(24)  & 1.8(7)  \\
\specialrule{0em}{1pt}{1pt}
$^{48}$Ni  &\cite{pomorski2014proton} &469$^{\text{a}}$ &1290(40) & $3.0_{-1.2}^{+2.2}$  & $1.61_{-1.04}^{+3.20}$  & $11.71_{-7.48}^{+22.26}$  & 1.3(6)  & 0.3(2)  \\
\specialrule{0em}{1pt}{1pt}
$^{54}$Zn  &\cite{blank2005first,ascher2011direct} &120\tnote{$^\text{b}$} &1480(20) & $1.98_{-0.41}^{+0.73}$  & $0.76_{-0.28}^{+0.46}$  & $0.19_{-0.07}^{+0.11}$  & 1.7(8)  & 0.91(42)  \\
\specialrule{0em}{1pt}{1pt}
$^{67}$Kr  &\cite{goigoux2016two} &160\tnote{$^\text{c}$} &1690(17) & 20(11)  & $99_{-32}^{+48}$  & $125_{-40}^{+60}$  & 820(380)  & 280(130)  \\
\bottomrule
\end{tabular}
}
\begin{tablenotes}
\footnotesize
\item[1] $^\text{a}$From Ref. \cite{brown1991diproton}.
\item[2] $^\text{b}$From Ref. \cite{brown2002proton}.
\item[3] $^\text{c}$From Ref. \cite{ormand1997mapping}.
\end{tablenotes}
\end{table*}
In addition, as shown in Appendix A.2, the tri-body mode is mainly from lower angular momentum. The higher orbital tri-body half-lives, even with a pure TNA assumption, are much larger than those from the three-body calculation \cite{grigorenko2003prospective, grigorenko2003two} and diproton model. The larger tri-body half-lives for the [$d^2$] and [$f^2$] configurations may indicate that the decay of these high-$\ell$ configurations is dominated by decay mechanisms that are not contained in the tri-body model, such as diproton decay, which tends to give a lower half-life limit for these high-$\ell$ configurations than both the tri-body and three-body calculations. It is more possible that a realistic decay of these high-$\ell$ configurations situate in between the two extreme cases of diproton and tri-body.

\subsection{\label{sec:level2}The case of $^{67}Kr$}
For the heaviest case of $^{67}$Kr, the experimental half-life is much smaller than available theoretical predictions, especially for the three-body model and its shell model correction. To explain this difference, Grigorenko and coworkers propose a mix of one- and two-proton decay mechanisms due to the small negative proton separation energy \cite{grigorenko2017decay}. Wang and Nazarewicz have studied the influence of deformation couplings on the $2p$ decay of $^{67}$Kr in the GCC model \cite{wang2018puzzling}. They found that the deformation effect can increase the $p$-wave component and thus decrease the $2p$ decay half-life.

With Eq.~(\ref{eq8}), we can evaluate the sensitivity of the resonance energy $E_r$ of the $1p$ daughter in $^{67}$Kr and the weight of [$p^2$] configuration to the direct $2p$ decay width. Results are presented in FIG. 2, where the vertical line of $Q_{2p}$ separates the energy range into two regions: for $E_r<Q_{2p}$, both sequential and direct $2p$ emissions are allowed, while for $E_r>Q_{2p}$, only direct $2p$ emission is energetically possible. The sequential width in the allowed energy region is calculated by Eq.~(\ref{eq3}) and~(\ref{eq4}), assuming a pure $p$-wave component. Similar calculations were also performed with the three-body model and improved direct decay model (IDDM) \cite{grigorenko2017decay}. Comparing these results shows that the $T_{1/2}$ values of the tri-body model are larger than those of the IDDM but close to those of the three-body model around $Q_{2p}$. In addition, tri-body results vary more slowly, especially for $E_r<1.4$ MeV, where other decay mechanisms, like the two-body decay, may become important and a quick drop of half-life is observed in their results. Despite these differences, the conclusion is similar: the half-lives decrease with small p-wave resonance energy $E_r$ in $^{66}$Br and large weights of [$p^2$] configuration. High purity of $p$-wave component can reproduce the experimental data within an extensive range of resonance energy across $Q_{2p}$. However, the shell model calculation predicts a [$p^2$] configuration of less than 20\% \cite{brown2019hybrid}. In this case, the experimental decay width will give resonance energy $E_r$ much smaller than the two-proton separation energy, where the sequential decay contribution may become more critical.

\begin{figure}[hbt]
\vspace{-0.4cm}
\setlength{\abovecaptionskip}{0cm}
\setlength{\belowcaptionskip}{-0.2cm}
\includegraphics[width=0.5\textwidth]{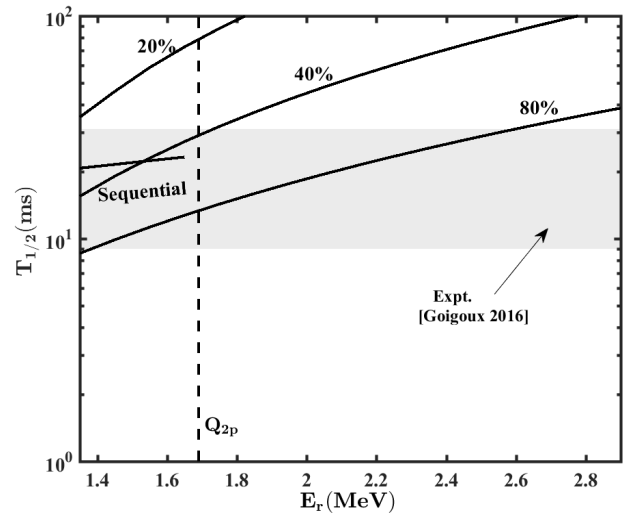}
\captionsetup{font={small}}
\caption{\label{fig:2} Two-proton decay half-life as a function of the resonance energy and weight of $[p^2]$ configuration}
\end{figure}

From FIG. 2, the sequential decay width is less sensitive to the resonance energy of the intermediate state and the predicted results agree with the experimental data in the allowed region. The calculated half-life of sequential decay, for $E_r$=1.38 MeV, is of ${20.80}_{-6.57}^{+9.77}$ ms (within experimental $2p$ decay energy uncertainty). Therefore, it is possible that with a high purity of the $p$-wave component due to the deformation, direct $2p$ decay or a mixture of sequential and direct $2p$ decay (depending on the energy level of the $1p$ daughter) can occur in $^{67}$Kr. Further verification relies on a more precise determination of nuclear mass for the intermediate state and experimental $p$-$p$ angular correlations measurement with more significant statistics.
\subsection{\label{sec:level3}Predicted half-life for unobserved $2p$ radioactivity}
\begin{table*}[ht]
\setlength{\abovecaptionskip}{0cm}
\setlength{\belowcaptionskip}{-0.2cm}
\captionsetup{font={normalsize},justification=raggedright}
\caption{\label{tab:table 2} Calculated half-lives (in ms) for different $2p$ radioactivity candidates in the region $22<Z<50$ using the diproton model of Eq.~(\ref{eq4}) with Barker's formulation \cite{barker200112} and the tri-body model of Eq.~(\ref{eq8}). All the calculations were made with $\ell_{2p}=0$ for the diproton case and $\ell_p=1$ for the tri-body case. The spectroscopic factors in both cases are assumed to be one.}
\resizebox{\linewidth}{!}{
\scriptsize
\begin{tabular}{cccccccccc}
\toprule
 & & &\multicolumn{3}{c}{Diproton model}&\multicolumn{3}{c}{Tri-body model}&\\
 \cmidrule(r){4-6}  \cmidrule(r){7-9}
\specialrule{0em}{1pt}{1pt}
$^{A}Z$ &$S_p$(MeV)  &$S_{2p}$(MeV)  &$T_{1/2}^{2p}$ &$T_{1/2,min}^{2p}$&$T_{1/2,max}^{2p}$ &$T_{1/2}^{2p}$ &$T_{1/2,min}^{2p}$&$T_{1/2,max}^{2p}$  &$T_{1/2}^{2p}$ in Ref.\cite{ormand1996properties,ormand1997mapping} \\ 
\midrule
\specialrule{0em}{1pt}{1pt}
$^{96}$Sn$^{\text{a}}$\tnote{$^{\text{a}}$}  & 0.29  & -1.78(27)   & $4.60\times 10^{9}$  & $3.62\times 10^{6}$  & $3.66\times 10^{13}$ & $8.12\times 10^{8}$  & $7.89\times 10^{5}$  & $4.88\times 10^{12}$  &-  \\
\specialrule{0em}{1pt}{1pt}
$^{92}$Cd$^{\text{a}}$  & 0.61  & -1.45(27)   & $1.15\times 10^{13}$  & $1,3\times 10^{9}$  & $1.88\times 10^{18}$ & $1.20\times 10^{12}$  & $1.81\times 10^{8}$  & $1.32\times 10^{17}$ &-  \\
\specialrule{0em}{1pt}{1pt}
$^{88}$Pd$^{\text{a}}$  & 0.59  & -1.74(27)   & $2.78\times 10^{7}$  & $3.39\times 10^{4}$  & $1.30\times 10^{11}$ & $4.95\times 10^{6}$  & $7.42\times 10^{3}$  & $1.76\times 10^{10}$ &-  \\
\specialrule{0em}{1pt}{1pt}
$^{84}$Ru$^{\text{a}}$  & 0.24  & -1.59(26)   & $1.05\times 10^{8}$  & $7.16\times 10^{4}$  & $1.25\times 10^{12}$ & $1.90\times 10^{7}$  & $1.65\times 10^{4}$  & $1.64\times 10^{11}$ &-  \\
\specialrule{0em}{1pt}{1pt}
$^{80}$Mo$^{\text{a}}$  & 0.48  & -1.06(26)   & $4.89\times 10^{15}$  & $2.79\times 10^{10}$  & $2.23\times 10^{23}$ & $3.39\times 10^{14}$  & $2.99\times 10^{9}$  & $7.94\times 10^{21}$ &-  \\
\specialrule{0em}{1pt}{1pt}
$^{74}$Zr$^{\text{a}}$  & 0.03  & -2.58(26)   & $6.13\times 10^{-4}$  & $2.33\times 10^{-5}$  & $2.80\times 10^{-2}$ & $4.07\times 10^{-4}$  & $1.68\times 10^{-5}$  & $1.67\times 10^{-2}$ &-  \\
\specialrule{0em}{1pt}{1pt}
$^{71}$Sr$^{\text{b}}$\tnote{$^{\text{b}}$}  & -0.02(15)  &-2.06(14)& $0.146$  & $1.33\times 10^{-2}$  & 2.08 & $7.07\times 10^{-2}$  & $6.94\times 10^{-3}$  & $0.92$ &-  \\
\specialrule{0em}{1pt}{1pt}
$^{63}$Se$^{\text{b}}$  & 0.11(14)  &-1.51(14)& $41.9$  & $1.45$  & $2.02\times 10^{3}$ & $13.2$  & $0.52$  & $5.46\times 10^{2}$ &$3\times 10^{-4}$-50  \\
\specialrule{0em}{1pt}{1pt}
$^{59}$Ge$^{\text{b}}$  & 0.19(14)  &-1.16(14)& $4.22\times 10^{4}$  & $417$  & $1.07\times 10^{7}$ & $8.49\times 10^{3}$  & $1.02\times 10^{2}$  & $1.69\times 10^{6}$ &0.01-300  \\
\specialrule{0em}{1pt}{1pt}
$^{42}$Cr$^{\text{c}}$\tnote{$^{\text{c}}$}  & 1.28(20)  &-0.45(15)& $8.45\times 10^{13}$  & $1.1\times 10^{8}$  & $8.25\times 10^{23}$ & $2.33\times 10^{12}$  & $8.18\times 10^{6}$  & $3.04\times 10^{21}$ &$10^5$-$10^{19}$  \\
\specialrule{0em}{1pt}{1pt}
$^{39}$Ti$^{\text{c}}$  & 0.48(13)  &-0.67(11) & $8.20\times 10^{3}$  & $38.4$  & $7.56\times 10^{6}$ & $1.24\times 10^{3}$  & $7.84$  & $7.74\times 10^{5}$ &0.4-2000  \\
\specialrule{0em}{1pt}{1pt}
$^{38}$Ti$^{\text{c}}$  & 0.44(16)  & -2.43(13)  & $3.21\times 10^{-12}$  & $1.19\times 10^{-12}$  & $9.46\times 10^{-12}$ & $5.08\times 10^{-12}$  & $1.85\times 10^{-12}$  & $1.50\times 10^{-11}$ &(0.4-2.3)$\times 10^{-12}$  \\
\bottomrule
\end{tabular}
}
\begin{tablenotes}
\footnotesize
\item[1]$^\text{a}$One- and two-proton separation energies from Ref. \cite{cai2022shell}.
\item[2]$^\text{b}$One- and two-proton separation energies from Ref. \cite{brown2002proton}.
\item[3]$^\text{c}$One- and two-proton separation energies from Ref. \cite{ormand1996properties}.
\end{tablenotes}
\end{table*}
The candidates for the ground state $2p$ radioactivity satisfy specific energy constraints of positive one-proton separation energy while negative two-proton separation energy due to the pairing force. Therefore, the theoretical prediction requires precise estimations of nuclear masses in the proton-rich region. Previous investigations have been carried out for the region $22<Z<30$ \cite{brown1991diproton,cole1996stability,ormand1996properties} and $30<Z<38$ \cite{brown2002proton,ormand1997mapping} based on shell model calculation with the isobaric mass multiplet equation (IMME). Other approaches have also been proposed for heavier regions. Recently, two analytical formulae with a systematic uncertainty of about 300 keV have been proposed for the region $30<Z<50$ \cite{cai2022shell}. The density functional theory achieved the extension to a heavier region above strontium \cite{olsen2013landscape}. By comparing with other competitive decay modes, such as $\beta^+$ decay, some potential $2p$ candidates have been predicted. This section employs the diproton and tri-body models with the predicted separation energies for the unobserved $2p$ radioactivity candidates in the region $22<Z<50$. The spectroscopic factors in both models are taken to be one for simplicity, which is expected to bring about one order of difference in the final prediction but does not influence the conclusion. In addition, all calculations were made with $\ell_{2p}=0$ for the diproton case (two protons in the $\ell=S=0$ state) and $\ell_p=1$ for the single-proton state in the tri-body model, which is also the fastest decay possible for protons released from $pf$ shells. The results are presented in Table 2, including the predicted one- and two-proton separation energies from different theoretical methods and the calculated half-lives for different $2p$ radioactivity candidates. The minimum and maximum values result from the uncertainties of two-proton separation energies. Moreover, previous half-life predictions for some nuclei from Ref. \cite{ormand1996properties,ormand1997mapping} are also listed in Table 2. 

The range of observable half-life for $2p$ emission is mainly determined by experimental conditions and competing decay mechanisms, such as $\beta^+$ decay, the half-life of which is on the order of 10-100 ms in this region \cite{ormand1996properties}. The lower limit for observable lifetime is determined by the sensitivity of experimental setups, such as 100 ns for the in-flight, projectile-fragmentation techniques \cite{Pfutzner_2013}. By taking these two limits for the observable lifetime, $^{71}$Sr and $^{74}$Zr may be the best candidates. The case of $^{63}$Se may fail to compete with $\beta^+$ decay with a more realistic spectroscopic factor. The case of $^{38}$Ti may be more difficult to be observed due to its much shorter half-life. Meanwhile, $^{71}$Sr has two more neutrons beyond the most neutron-deficient isotope known to date, while $^{74}$Zr has four more neutrons than the current experimental reach. Therefore, $^{71}$Sr may be easier to reach experimentally and is promising for further research. However, from Table 2, a large uncertainty of the calculated $T_{1/2}$ resulting from the uncertainty of $S_{2p}$ is also observed. The conclusion drawn above may be changed by a more precise $S_{2p}$ measurement in future experiments.

\subsection{\label{sec:level4}The case of $^{18}$Mg}
Due to a weak Coulomb barrier in the light-mass region, some unbound proton-rich nuclei exhibit much shorter half-lives with more exotic decay patterns. For instance, $^{18}$Mg is predicted to be unbound by one-, two-, and even four-proton emissions and was recently observed to decay by two sequential steps of direct $2p$ emission. The measured decay energy and the decay width are 4.865(34) MeV and 115(100) keV, respectively \cite{jin2021first}, which are in reasonable agreement with previous predictions in the Gamow shell model (GSM) \cite{michel2021proton}. They calculated the decay energy of 4.898 MeV and a total decay width of 98 keV. They have also used this method to calculate the $2p$ decay width of about 10-15 keV. In this section, we attempt to estimate the two- and four-proton decay widths in the ground state of $^{18}$Mg with the present derived formulae. 

For the case of ground state $2p$ emission in $^{18}$Mg, initial and final wave functions are calculated with the modified YSOX Hamiltonian \cite{yuan2012shell} in the model space of four protons in the $1s$-$0d$ shells and two neutron holes in the $0p$ shell. The calculated $\theta_{2p}^2$ are listed in Appendix A.1. Therefore, with a $2p$ decay energy of 3.47 MeV, i.e., the difference between the experimental masses of $^{18}$Mg and $^{16}$Ne \cite{mukha2008proton}, the calculated $2p$ decay width is 29 keV, slightly higher than the prediction given by GSM.

For the case of sequential $4p$ emission, without considering the internal energies of the two emitted diprotons, one can directly apply the sequential formulae of Eq.~(\ref{eq3}) and~(\ref{eq4}) to estimate the decay width. To calculate $\theta_{2p}^2$ for the second step of diproton emission, YSOX Hamiltonian is applied to $^{16}$Ne with two protons in the $1s$-$0d$ shells and two neutron holes in the $0p$ shell. Finally, by combining the spectroscopic factors information and the decay energies of the parent and intermediate nuclei, the calculated width is 55(4) keV (with uncertainty from the measured $Q_{4p}$). The result agrees with experimental data of 115(100) keV. With an adequate hypothesis, including internal energy in the emitted diprotons may decrease the penetration energy and thus reduce the decay width. Eq.~(\ref{eq9}) is used to quantify this effect, and the resulting width decreases to 1 keV. Therefore, the value of 55(4) keV can be regarded as an upper limit for the sequential $4p$ emission in $^{18}$Mg. 
\section{\label{sec:level5}Conclusion}
In this letter, we propose, from the extended R-matrix theory, decay width formulae for different $2p$ emission mechanisms, including sequential $2p$ decay, diproton decay, and tri-body decay. Combined with spectroscopic factors obtained through shell-model calculation, these formulae are used to investigate half-lives for the currently observed five ground state $2p$ radioactivity cases. The diproton and tri-body formulae can reproduce experimental $2p$ half-lives well. Moreover, considering the applicability, our formula is promising for future $2p$ decay half-life estimation.

For the case of $^{67}$Kr, a mix of sequential and direct contributions are analyzed, including the small negative separation energy and a sizeable $p$-wave component. Further verification requires more precise measurements of intermediate state mass and the $p$-$p$ angular correlation. In addition, more $2p$ candidates are also proposed in the region $22<Z<50$ using the diproton and tri-body models. It is concluded that the most promising candidates for $2p$ radioactivity are $^{71}$Sr and $^{74}$Zr. $^{38}$Ti presents a much shorter half-life and may be difficult to be observed with current techniques. The conclusion may be updated by future experimental progress due to a large $S_{2p}$ sensitivity. Finally, the sequential decay width formula for the ground state sequential $4p$ emission in $^{18}$Mg gives an upper limit of 55(4) keV, which agrees with the previous measurement of 115(100) keV.

\section*{Acknowledgments}
Discussions with Jiajian Liu are greatly acknowledged. This work has been supported by the Guangdong Major Project of Basic and Applied Basic Research under Grant No. 2021B0301030006, the National Natural Science Foundation of China under Grant Nos. 11825504, 11961141004, the computational resources from SYSU and National Supercomputer Center in Guangzhou.

\appendix
\section{}
\subsection{}
\setcounter{table}{0}
\begin{table}[!h]
\setlength{\abovecaptionskip}{0cm}
\setlength{\belowcaptionskip}{-0.2cm}
\caption{Calculated two-proton spectroscopic factors $\theta_{2p}^2$ for different nuclei through Eq.~(\ref{eq10}) with TNAs from Ref. \cite{brown2019hybrid} ($^{19}$Mg, $^{45}$Fe, $^{48}$Ni, $^{54}$Zn, $^{67}$Kr) and YSOX Hamiltonian calculation ($^{16}$Ne, $^{18}$Mg).}
\resizebox{\linewidth}{!}{
\begin{tabular}{ccccccccc}
\toprule
\specialrule{0em}{1pt}{1pt}
 Orbit&$^{16}$Ne &$^{18}$Mg&$^{19}$Mg  &Orbit &$^{45}$Fe &$^{48}$Ni &$^{54}$Zn  &$^{67}$Kr \\ 
\midrule
\specialrule{0em}{1pt}{1pt}
$0d_{5/2}$&0.197  &0.327 & 0.374  & $0f_{7/2}$  & 0.256  & 0.226  & 0.010  & 0.003\\
\specialrule{0em}{1pt}{1pt}
$0d_{3/2}$&0.015  &0.026 & 0.026  & $0f_{5/2}$  & 0.003  & 0.002  & 0.008  & 0.135\\
\specialrule{0em}{1pt}{1pt}
$1s_{1/2}$&0.225  &0.146 & 0.105  & $1p_{3/2}$  & 0.013  & 0.007  & 0.131  & 0.039\\
\specialrule{0em}{1pt}{1pt}
          &  & &        & $1p_{1/2}$  & 0.001  & 0.001  & 0.008  & 0.018\\
\specialrule{0em}{1pt}{1pt}
          &  & &        & $1s_{1/2}$  & 0.011  & 0.009  & 0.002  & 0.001\\
\bottomrule        
\end{tabular}
}
\end{table}

\subsection{}
\begin{table}[!h]
\setlength{\abovecaptionskip}{0cm}
\setlength{\belowcaptionskip}{-0.2cm}
\caption{Tri-body, diproton, and three-body half-lives (in ms) for different nuclei with a pure $\ell^2$ assumption. The results are based on experimental $Q_{2p}$ values from Table 1. The tri-body and diproton half-lives are calculated by Eq.~(\ref{eq8}) and Eq.~(\ref{eq4}), respectively, with TNA($j^2$)=1. The three-body half-lives are taken from Ref. \cite{grigorenko2003prospective} for $^{19}$Mg and Ref. \cite{grigorenko2003two} for $^{45}$Fe, $^{48}$Ni, $^{54}$Zn and $^{67}$Kr. The three-body $s^2$ half-lives are estimated from Ref. \cite{brown2019hybrid}.}
\resizebox{\linewidth}{!}{
\begin{tabular}{ccccccccc}
\toprule
\specialrule{0em}{1pt}{1pt}
  &Orbit &$^{19}$Mg&$^{45}$Fe &$^{48}$Ni &$^{54}$Zn  &$^{67}$Kr \\ 
\midrule
\specialrule{0em}{1pt}{1pt}
Diproton $T_{1/2}$&$1s^2$  &$2.09_{-1.44}^{+5.53}$ ps & $0.77_{-0.28}^{+0.46}$  & $0.84_{-0.54}^{+1.66}$  & 0.23(11)  & 32(10)\\
\specialrule{0em}{1pt}{1pt}
                  &$1p^2$  &- & 1.3(6)  & $1.35_{-0.88}^{+2.69}$  & $0.39_{-0.14}^{+0.23}$  & 54(20)\\
\specialrule{0em}{1pt}{1pt}
                  &$0d^2$  & $3.48_{-2.41}^{+9.26}$ ps & -  & -  & -  & -\\
\specialrule{0em}{1pt}{1pt}
                  &$0f^2$  &- & $1.49_{-0.55}^{+0.9}$  & $1.59_{-1.03}^{+3.15}$  & $0.59_{-0.22}^{+0.36}$  & $65(25)$\\
\specialrule{0em}{1pt}{1pt}
\\
Tri-body $T_{1/2}$&$1s^2$  &$0.11_{-0.07}^{+0.25}$ ps & 0.03(1)  & $0.05_{-0.03}^{+0.1}$  & 0.014(8)  & 2.1(7)\\
\specialrule{0em}{1pt}{1pt}
                  &$1p^2$  &- & 0.25(9)  & $0.4_{-0.3}^{+0.8}$  & 0.10(5)  & 12(4)\\
\specialrule{0em}{1pt}{1pt}
                  &$0d^2$  & $ \approx10^3$ ps & -  & -  & -  & -\\
\specialrule{0em}{1pt}{1pt}
                  &$0f^2$  &- & $10^3 - 10^4$  & $10^3 - 10^4$  & $10^2 - 10^3$  & $\approx10^4$\\
\specialrule{0em}{1pt}{1pt}
\\
Three-body $T_{1/2}$&$1s^2$  &$0.2_{-0.05}^{+0.37}$ ps & 0.24(12)  & 0.04(3)  & 0.12(6)  & 38(18)\\
\specialrule{0em}{1pt}{1pt}
                  &$1p^2$  &-& 1.8(7)  & 0.3(2)  & 0.91(42)  & 280(130)\\
\specialrule{0em}{1pt}{1pt}
                  &$0d^2$  &$2.5_{-0.5}^{+6.5}$ ps & -  & -  & -  & -\\
\specialrule{0em}{1pt}{1pt}
                  &$0f^2$  &-& 99(40)  & 10(5)  & 45(27)  &  $\approx10^4$ \\
\bottomrule        
\end{tabular}
}
\end{table}

\biboptions{numbers,sort&compress}
\bibliographystyle{elsarticle-num} 
\bibliography{references}

\begin{thebibliography}{10}
\expandafter\ifx\csname url\endcsname\relax
  \def\url#1{\texttt{#1}}\fi
\expandafter\ifx\csname urlprefix\endcsname\relax\def\urlprefix{URL }\fi
\expandafter\ifx\csname href\endcsname\relax
  \def\href#1#2{#2} \def\path#1{#1}\fi

\bibitem{goldansky1961two}
V.~Goldansky, Two-proton radioactivity, Nuclear Physics 27~(4) (1961) 648--664.

\bibitem{pfutzner2002first}
M.~Pf{\"u}tzner, E.~Badura, C.~Bingham, B.~Blank, M.~Chartier, H.~Geissel,
  J.~Giovinazzo, L.~Grigorenko, R.~Grzywacz, M.~Hellstr{\"o}m, et~al., First
  evidence for the two-proton decay of $^{45}${Fe}, The European Physical
  Journal A-Hadrons and Nuclei 14~(3) (2002) 279--285.

\bibitem{zhou2022}
L.~Zhou, S.~M. Wang, D.~Q. Fang, Y.~G. Ma, Recent progress in two-proton
  radioactivity, Nuclear Science and Techniques 33~(8) (2022) 105.

\bibitem{giovinazzo2002two}
J.~Giovinazzo, B.~Blank, M.~Chartier, S.~Czajkowski, A.~Fleury, M.~J.
  LopezJimenez, M.~S. Pravikoff, J.-C. Thomas, F.~de~Oliveira~Santos,
  M.~Lewitowicz, et~al., Two-proton radioactivity of $^{45}${Fe}, Physical
  review letters 89~(10) (2002) 102501.

\bibitem{blank2005first}
B.~Blank, A.~Bey, G.~Canchel, C.~Dossat, A.~Fleury, J.~Giovinazzo, I.~Matea,
  N.~Adimi, F.~De~Oliveira, I.~Stefan, et~al., First observation of $^{54}${Zn}
  and its decay by two-proton emission, Physical review letters 94~(23) (2005)
  232501.

\bibitem{dossat2005two}
C.~Dossat, A.~Bey, B.~Blank, G.~Canchel, A.~Fleury, J.~Giovinazzo, I.~Matea,
  F.~de~Oliveira~Santos, G.~Georgiev, S.~Gr{\'e}vy, et~al., Two-proton
  radioactivity studies with $^{45}${Fe} and $^{48}${Ni}, Physical Review C
  72~(5) (2005) 054315.

\bibitem{mukha2007observation}
I.~Mukha, K.~S{\"u}mmerer, L.~Acosta, M.~A.~G. Alvarez, E.~Casarejos,
  A.~Chatillon, D.~Cortina-Gil, J.~Espino, A.~Fomichev, J.~E. Garcia-Ramos,
  et~al., Observation of two-proton radioactivity of $^{19}${Mg} by tracking
  the decay products, Physical review letters 99~(18) (2007) 182501.

\bibitem{goigoux2016two}
T.~Goigoux, P.~Ascher, B.~Blank, M.~Gerbaux, J.~Giovinazzo, S.~Gr{\'e}vy, T.~K.
  Nieto, C.~Magron, P.~Doornenbal, G.~Kiss, et~al., Two-proton radioactivity of
  $^{67}${Kr}, Physical review letters 117~(16) (2016) 162501.

\bibitem{olsen2013landscape}
E.~Olsen, M.~Pf{\"u}tzner, N.~Birge, M.~Brown, W.~Nazarewicz, A.~Perhac,
  Landscape of two-proton radioactivity, Physical Review Letters 110~(22)
  (2013) 222501.

\bibitem{grigorenko2000theory}
L.~V. Grigorenko, R.~C. Johnson, I.~G. Mukha, I.~J. Thompson, M.~V. Zhukov,
  Theory of two-proton radioactivity with application to $^{19}${Mg} and
  $^{48}${Ni}, Physical Review Letters 85~(1) (2000) 22.

\bibitem{grigorenko2001two}
L.~V. Grigorenko, R.~C. Johnson, I.~G. Mukha, I.~J. Thompson, M.~V. Zhukov,
  Two-proton radioactivity and three-body decay: General problems and
  theoretical approach, Physical Review C 64~(5) (2001) 054002.

\bibitem{barker1999width}
F.~C. Barker, Width of the $^{12}${O} ground state, Physical Review C 59~(1)
  (1999) 535.

\bibitem{barker200112}
F.~C. Barker, $^{12}${O} ground-state decay by $^2${He} emission, Physical
  Review C 63~(4) (2001) 047303.

\bibitem{brown2003di}
B.~A. Brown, F.~C. Barker, Di-proton decay of $^{45}${Fe}, Physical Review C
  67~(4) (2003) 041304.

\bibitem{brown2002di}
B.~A. Brown, F.~C. Barker, D.~J. Millener, Di-proton decay of the 6.15 {MeV}
  $1^-$ state in $^{18}${Ne}, Physical Review C 65~(5) (2002) 051309.

\bibitem{lane1958r}
A.~Lane, R.~Thomas, R-matrix theory of nuclear reactions, Reviews of Modern
  Physics 30~(2) (1958) 257.

\bibitem{charity20102}
R.~J. Charity, J.~M. Elson, J.~Manfredi, R.~Shane, L.~G. Sobotka, Z.~Chajecki,
  D.~Coupland, H.~Iwasaki, M.~Kilburn, J.~Lee, et~al., 2p-2p decay of $^{8}${C}
  and isospin-allowed 2p decay of the isobaric-analog state in $^{8}${B},
  Physical Review C 82~(4) (2010) 041304.

\bibitem{jin2021first}
Y.~Jin, C.~Y. Niu, K.~W. Brown, Z.~H. Li, H.~Hua, A.~K. Anthony, J.~Barney,
  R.~J. Charity, J.~Crosby, D.~Dell’Aquila, et~al., First observation of the
  four-proton unbound nucleus $^{18}${Mg}, Physical Review Letters 127~(26)
  (2021) 262502.

\bibitem{teichmann1952sum}
T.~Teichmann, E.~P. Wigner, Sum rules in the dispersion theory of nuclear
  reactions, Physical Review 87~(1) (1952) 123.

\bibitem{anyas1974nuclear}
N.~Anyas-Weiss, J.~Cornell, P.~Fisher, P.~Hudson, A.~Menchaca-Rocha,
  D.~Millener, A.~Panagiotou, D.~Scott, D.~Strottman, D.~Brink, et~al., Nuclear
  structure of light nuclei using the selectivity of high energy transfer
  reactions with heavy ions, Physics Reports 12~(3) (1974) 201--272.

\bibitem{audirac2012direct}
L.~Audirac, P.~Ascher, B.~Blank, C.~Borcea, B.~Brown, G.~Canchel, C.~Demonchy,
  F.~de~Oliveira~Santos, C.~Dossat, J.~Giovinazzo, et~al., Direct and
  $\beta$-delayed multi-proton emission from atomic nuclei with a time
  projection chamber: the cases of $^{43}${Cr}, $^{45}${Fe}, and $^{51}${Ni},
  The European Physical Journal A 48~(12) (2012) 1--12.

\bibitem{brown2019hybrid}
B.~A. Brown, B.~Blank, J.~Giovinazzo, Hybrid model for two-proton
  radioactivity, Physical Review C 100~(5) (2019) 054332.

\bibitem{grigorenko2003prospective}
L.~V. Grigorenko, I.~G. Mukha, M.~V. Zhukov, Prospective candidates for the
  two-proton decay studies {I}: structure and coulomb energies of $^{17}${Ne}
  and $^{19}${Mg}, Nuclear Physics A 713~(3-4) (2003) 372--389.

\bibitem{grigorenko2003two}
L.~V. Grigorenko, M.~V. Zhukov, Two-proton radioactivity and three-body decay.
  {II}. exploratory studies of lifetimes and correlations, Physical Review C
  68~(5) (2003) 054005.

\bibitem{mukha2008proton}
I.~Mukha, L.~Grigorenko, K.~S{\"u}mmerer, L.~Acosta, M.~Alvarez, E.~Casarejos,
  A.~Chatillon, D.~Cortina-Gil, J.~M. Espino, A.~Fomichev, et~al.,
  Proton-proton correlations observed in two-proton decay of $^{19}${Mg} and
  $^{16}${Ne}, Physical Review C 77~(6) (2008) 061303.

\bibitem{ascher2011direct}
P.~Ascher, L.~Audirac, N.~Adimi, B.~Blank, C.~Borcea, B.~A. Brown, I.~Companis,
  F.~Delalee, C.~E. Demonchy, F.~de~Oliveira~Santos, et~al., Direct observation
  of two protons in the decay of $^{54}${Zn}, Physical Review Letters 107~(10)
  (2011) 102502.

\bibitem{miernik2007two}
K.~Miernik, W.~Dominik, Z.~Janas, M.~Pf{\"u}tzner, L.~Grigorenko, C.~R.
  Bingham, H.~Czyrkowski, M.~{\'C}wiok, I.~G. Darby, R.~D{\c{a}}browski,
  et~al., Two-proton correlations in the decay of $^{45}${Fe}, Physical Review
  Letters 99~(19) (2007) 192501.

\bibitem{pomorski2014proton}
M.~Pomorski, M.~Pf{\"u}tzner, W.~Dominik, R.~Grzywacz, A.~Stolz, T.~Baumann,
  J.~S. Berryman, H.~Czyrkowski, R.~D{\c{a}}browski, A.~Fija{\l}kowska, et~al.,
  Proton spectroscopy of $^{48}${Ni}, $^{46}${Fe}, and $^{44}${Cr}, Physical
  Review C 90~(1) (2014) 014311.

\bibitem{brown1991diproton}
B.~A. Brown, Diproton decay of nuclei on the proton drip line, Physical Review
  C 43~(4) (1991) R1513.

\bibitem{brown2002proton}
B.~A. Brown, R.~R.~C. Clement, H.~Schatz, A.~Volya, W.~A. Richter, Proton
  drip-line calculations and the $rp$ process, Physical Review C 65~(4) (2002)
  045802.

\bibitem{ormand1997mapping}
W.~E. Ormand, Mapping the proton drip line up to {A}=70, Physical Review C
  55~(5) (1997) 2407.

\bibitem{grigorenko2017decay}
L.~V. Grigorenko, T.~A. Golubkova, J.~S. Vaagen, M.~V. Zhukov, Decay mechanism
  and lifetime of $^{67}${Kr}, Physical Review C 95~(2) (2017) 021601.

\bibitem{wang2018puzzling}
S.~M. Wang, W.~Nazarewicz, et~al., Puzzling two-proton decay of $^{67}${Kr},
  Physical Review Letters 120~(21) (2018) 212502.

\bibitem{ormand1996properties}
W.~E. Ormand, Properties of proton drip-line nuclei at the $sd$-$fp$-shell
  interface, Physical Review C 53~(1) (1996) 214.

\bibitem{cai2022shell}
B.~Cai, G.~Chen, C.~Yuan, J.~He, Shell-model study on properties of proton
  dripline nuclides with {Z, N}=30-50 including uncertainty analysis, Chinese
  Physics C (2022).

\bibitem{cole1996stability}
B.~Cole, Stability of proton-rich nuclei in the upper $sd$ shell and lower $pf$
  shell, Physical Review C 54~(3) (1996) 1240.

\bibitem{Pfutzner_2013}
M.~Pfutzner, Particle radioactivity of exotic nuclei, Physica Scripta
  2013~(T152) (2013) 014014.

\bibitem{michel2021proton}
N.~Michel, J.~G. Li, F.~R. Xu, W.~Zuo, Proton decays in $^{16}${Ne} and
  $^{18}${Mg} and isospin-symmetry breaking in carbon isotopes and isotones,
  Physical Review C 103~(4) (2021) 044319.

\bibitem{yuan2012shell}
C.~Yuan, T.~Suzuki, T.~Otsuka, F.~Xu, N.~Tsunoda, Shell-model study of boron,
  carbon, nitrogen, and oxygen isotopes with a monopole-based universal
  interaction, Physical Review C 85~(6) (2012) 064324.

\end{thebibliography}

\end{document}